\documentclass{styles/svproc}
\usepackage{url}
\usepackage{graphicx}
\usepackage{float}
\usepackage{lipsum}

\usepackage{tikz,xcolor,hyperref}

\definecolor{lime}{HTML}{A6CE39}
\DeclareRobustCommand{\orcidicon}{%
	\begin{tikzpicture}
	\draw[lime, fill=lime] (0,0) 
	circle [radius=0.16] 
	node[white] {{\fontfamily{qag}\selectfont \tiny ID}};
	\draw[white, fill=white] (-0.0625,0.095) 
	circle [radius=0.007];
	\end{tikzpicture}
	\hspace{-2mm}
}

\foreach \x in {A, ...,Z}{%
	\expandafter\xdef\csname orcid\x\endcsname{\noexpand\href{https://orcid.org/\csname orcidauthor\x\endcsname}{\noexpand\orcidicon}}
}

\begin{document}
\mainmatter              
\title{DuoSearch: A Novel Search Engine for Bulgarian Historical Documents}
\titlerunning{DuoSearch}  
%
\author{Angel Beshirov\inst{1}\orcidA{}
\and Suzan Hadzhieva\inst{1}\orcidB{}
\and Ivan Koychev\inst{1}\orcidC{}
\and Milena Dobreva\inst{1}\orcidD{}}
\authorrunning{A. Beshirov et al.} 
%
\tocauthor{Angel Beshirov, Suzan Hadzhieva, Ivan Koychev, Milena Dobreva}
\institute{Faculty of Mathematics and Informatics, Sofia University “St. Kliment Ohridski”, Bulgaria}

\maketitle              

\begin{abstract}
Search in collections of digitised historical documents is hindered by a two-prong problem, orthographic variety and optical character recognition (OCR) mistakes. We present a new search engine for historical documents, DuoSearch, which uses ElasticSearch and machine learning methods based on deep neural networks to offer a solution to this problem. It was tested on a collection of historical newspapers in Bulgarian from the mid-19th to the mid-20th century. The system provides an interactive and intuitive interface for the end-users allowing them to enter search terms in modern Bulgarian and search across historical spellings. This is the first solution facilitating the use of digitised historical documents in Bulgarian. 
\end{abstract}
\section{Introduction and Related Work}
Many libraries and archives digitised sizeable collections and made an additional step towards datafication applying Optical Character Recognition (OCR).
However, digitised information is not easily accessible by end-users due to two main hindering blocks. First, the OCR process introduces errors in recognition because of challenging layouts and orthographic variety due to the nine language reforms applied to the Bulgarian language. Second, because the collections of historical documents include texts in a mixture of orthographic conventions the users should be able to use the historical forms of the search keywords. We are applying a novel approach that builds upon the automated techniques for post-OCR text correction in combination with spelling conversion and extended searching to tackle both issues (OCR errors and orthographic variety). Our search engine was used for a case study with a historical newspaper collection from The National Library "Ivan Vazov" (NLIV) in Plovdiv. 

The purpose of our research was to build a prototype search engine which addresses the two issues mentioned above. DuoSearch can be used for all kinds of digitised historical Bulgarian documents within the same period where OCR was not followed by quality control. The tool uses dictionary for Bulgarian but can be modified and adapted for other languages. 

The paper would be useful for anyone who is developing search tools for historical collections of texts with errors and/or linguistic variance. The system can perform a search in a collection of documents that are written in different spellings and have a relatively high number of erroneous tokens. It was implemented using ElasticSearch and uses machine learning techniques to improve the quality of the indexed data. It also provides an intuitive and interactive web interface, export of the results and easy navigation across the returned documents.

The interest in developing digital resources and tools which answer historians' needs for searching across collections of newspapers developed over the last decades \cite{allen}. Some researchers focused on newspaper collections metadata as a solution for improved search \cite{cole}. Many challenges arise when working with cultural heritage, like historical newspapers, some of which are described in \cite{oberbichler}. Common interests around these challenges in different disciplines have led to projects such as the European project NewsEye \cite{news:eye}. It uses data in a few languages and provides enhanced access to historical newspapers for a wide range of users. In Bulgaria, there is already a collection of digitised historical newspapers, but access to it is cumbersome, due to OCR errors and the multiple language reforms. There was a competition \cite{icdar} supported by NewsEye whose purpose was to compare and evaluate automatic approaches for correcting OCR-ed texts from historical documents. The competition provides data in 10 European languages including Bulgarian.

\section{System Design}
\begin{figure}
    \centering
	\includegraphics[width=0.7\columnwidth]{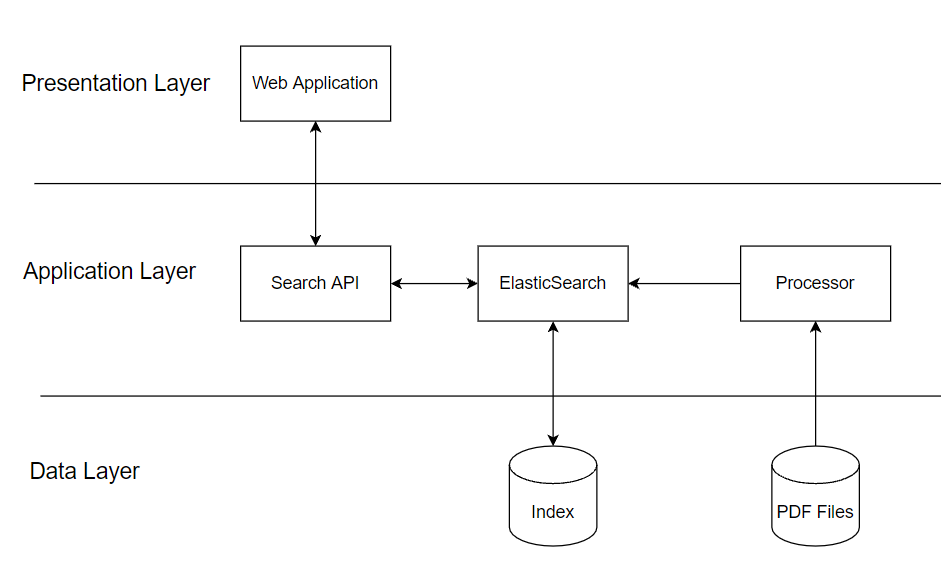}
	\caption{\label{fig:arch} 
	    System architecture
	}
\end{figure}

The design of the system is shown on figure \ref{fig:arch}. The system uses a three-tier architecture, where we have the user interface, the business logic and the data as independent modules. The presentation layer is responsible for the direct interaction between the user and the system, sending requests to the system and visualizing results. The search API component transforms and proxies the request to and from ElasticSearch in order to retrieve the relevant documents for the search query. To handle the linguistic variance, the Search API component contains a converter, which transforms the text entered by the user into the historical spelling and sends two requests.

The processor component does the data preprocessing before sending it to ElasticSearch for indexing. The preprocessing includes correcting mistakes from the post-OCRed text and removing redundant metadata from the pages. The post-OCR text correction process can be divided into two phases: error detection and error correction. For error detection we have used pretrained multilingual model BERT \cite{bert} which is context-aware and helps in identifying both syntactical and grammatical mistakes. The data is first tokenized into words, which are then transformed into BERT sub-tokens. Afterwards, we feed these subtokens to the BERT model and the output for each sub-token is passed to 5 convolutional layers with different kernel sizes: 2, 3, 4, 5 and 6, each one with 32 filters. After the convolutional layers, we have maxpool layers with a
stride of 1, which intuitively represents the information exchange between the n-grams. The outputs from these layers are then concatenated and fed to the last layer, which is linear and does the final classification of the sub-token. If more than one BERT sub-token is predicted to be erroneous, then the whole token is erroneous.

For error correction we use character-level sequence to sequence model which takes the classified as erroneous tokens from the error detection model and corrects them. For further improvement over the text correction we have used the CLaDA-BG dictionary \cite{clada} which contains over 1.1 million unique words in all of their correct forms. To address different orthographic varieties for post-OCR text correction, we need to know the version of the language in which the document is written, then we apply converter of the dictionary and use it in our model.

DuoSearch supports two search types: regular search and extended search. The regular search is a full text search using the standard match query from ElasticSearch. The extended search supports additional options, like proximity searching, boolean operators support and others. It is implemented using the query DSL of ElasticSearch.

For testing the current search engine prototype we have used a set of Bulgarian newspapers provided by NLIV, which is around 4 GB from the period 1882-1930. The source code is available on GitHub\footnote{\url{https://github.com/angelbeshirov/DuoSearch}} and live demo version on AWS.\footnote{\url{http://ec2-18-225-9-151.us-east-2.compute.amazonaws.com:8080}}.

\section{Evaluation}
We have used the Bulgarian dataset provided by the organizers of the ICDAR 2019 competition \cite{icdar} to evaluate our post-OCR text correction model. It contains 200 files with around 32,000 unique words from the period before 1945. For evaluation metrics of our text correction model we have used F-measure and \% of improvement. Our results are shown in Table 1 compared with the Clova AI model.

Our model is similar to the one of the Clova AI team from the competition with some improvements for the Bulgarian dataset. We managed to achieve an improvement of 9\% over it, using the additional dictionary  \cite{clada}. The \% of improvement is measured by comparing the weighted sum of the Levenshtein distances between the raw OCR text and the corrected one with the gold standard. We used the trained model afterwards to correct the mistakes from the PDF documents before indexing by the search engine. In future, additional evaluation will be done with the users of the system.

\begin{table}
\caption{Evaluation results}
\begin{center}
\begin{tabular}{c c c c c}
\hline
Model  & F-score & \% of improvement \\
\hline\rule{0pt}{14pt}
Clova AI & 0.77 & 9\% \\
DuoSearch & 0.79 & 18.7\% \\
\hline
\end{tabular}
\end{center}
\end{table}

\section{Conclusion and Contributions}
In this paper, we have described a new search engine that combines various technologies to allow for fast searching across a collection of historical newspapers. This contributes to the overall problem in Bulgaria with improving the access to historical documents. It has been acknowledged by Europeana as an example of successful partnership between universities and libraries.

In future, we will work on the text correction by developing a model which predicts the language revision version in which the text is written and apply different models for each orthography, as some of the words are being flagged as incorrect when they are written in different language version from the one the model is trained on. The system also has to be installed on the servers in the library and index the whole collection of newspapers, which is around 200 GB.

\section*{Acknowledgements}
This research is partially supported by Project UNITe BG05M2OP001-1.001-0004 funded by the OP ``Science and Education for Smart Growth'' and co-funded by the EU through the ESI Funds. The contribution of M. Dobreva is supported by the project KP-06-DB/6 DISTILL funded by the NSF of Bulgaria.

\end{document}